\documentclass{article}
\usepackage{arxiv}
\usepackage[utf8]{inputenc} 
\usepackage[T1]{fontenc}    
\usepackage{hyperref}       
\usepackage{url}            
\usepackage{booktabs}       
\usepackage{amsfonts}       
\usepackage{nicefrac}       
\usepackage{microtype}      
\usepackage{lipsum}		
\usepackage[numbers]{natbib}
\usepackage{doi}
\usepackage{graphicx}
\usepackage{caption}
\usepackage{subcaption}
\usepackage{array}
\newcolumntype{K}[1]{>{\centering\arraybackslash}p{#1}}
\usepackage{color}
\usepackage{wrapfig}
\usepackage{blindtext}
\usepackage{titlesec}

\usepackage{relsize}
\usepackage[printonlyused,withpage,smaller]{acronym}

\titleformat{\paragraph}
{\normalfont\normalsize\bfseries}{\theparagraph}{1em}{}
\titlespacing*{\paragraph}
{0pt}{3.25ex plus 1ex minus .2ex}{1.5ex plus .2ex}

\title{The World of Generative AI: Deepfakes and Large Language Models}

\author{
\href{https://orcid.org/0000-0002-8796-4819}{\includegraphics[scale=0.06]{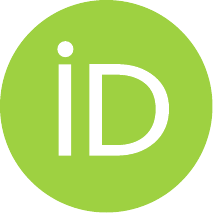}\hspace{1mm}Alakananda Mitra} \\ 
Nebraska Water Center\\ 
University of Nebraska-Lincoln, USA \\
\texttt{amitra6@unl.edu} \\
\And
\href{https://orcid.org/0000-0003-2959-6541}{\includegraphics[scale=0.06]{orcid.pdf}\hspace{1mm}Saraju P. Mohanty} \\
Dept. of Computer Science and Engineering \\
University of North Texas, USA \\
\texttt{ saraju.mohanty@unt.edu} \\
\And
\href{https://orcid.org/0000-0002-1616-7628}{\includegraphics[scale=0.06]{orcid.pdf}\hspace{1mm}Elias Kougianos} \\
Dept. of Electrical Engineering\\
University of North Texas, USA \\
\texttt{elias.kougianos@unt.edu} \\
}

\begin{document}

\maketitle

\begin{abstract}
We live in the era of Generative Artificial Intelligence (GenAI). Deepfakes and Large Language Models (LLMs) are two examples of GenAI. Deepfakes, in particular, pose an alarming threat to society as they are capable of spreading misinformation and changing the truth. LLMs are powerful language models that generate general-purpose language. However due to its generative aspect, it can also be a risk for people if used with ill intentions. The ethical use of these technologies is a big concern. This short article tries to find out the interrelationship between them.
\end{abstract}


\section{Introduction}
\label{Sec:Introduction}

The latest development in artificial intelligence (AI), chatbots, the product of generative AI, has captivated the public in the last two years. But it similarly poses an unprecedented challenge and can have potentially unwanted effects on our lives. OpenAI released the chatbot ChatGPT on November 30, 2022. The overwhelming response of the public towards ChatGPT usage pushed Google to release Bard, ChatGPT's rival, and Microsoft to release AI-powered Bing. But the recent GPT-4 topped the list as it has more capabilities than any other existing chatbot. Being LLM-based, these chatbots create synthetic media with the intention of creating better content, enhanced quality, or professional voices. The capabilities of such chatbots raise questions on the ethical use of AI.

In the meantime, deepfakes, which are high-quality AI-generated fake videos, have been circulating online. Synthetically generated deepfake videos have exceeded acceptable limits in terms of reality distortion. This disruptive technological development significantly impacts the truth \cite{mitra2021machine}. Due to the easy accessibility of generative AI models, the probability of misuse of this technology aggressively increases \cite{Shoaib2023DeepfakesMA}. Though deepfake technology came into this game in 2017, large language models like ChatGPT, Bing, Bard, GPT-4, etc. add a whole new dimension to this scenario. In particular, political figures are targeted in deepfake videos, contributing to the erosion of media credibility.


2024 is going to be an eventful year. A total of 78 countries are having major elections in 2024, according to the nonprofit organization for improving social media, the Integrity Institute \cite{IEEE_Spectrum}. Some examples are the U.S. presidential election, and national elections in South Africa, and the biggest democracy in the world, India. The extraordinary advent of generative AI, especially deepfake, has the potential of making these elections highly controversial \cite{IEEE_Spectrum}.

Deepfake videos are also showing up in Hollywood. Deceased actors can talk and even act in full-length feature films from beyond the grave, raising concerns over who holds the intellectual rights to a deceased person's semblance \cite{holldeep}. The most disturbing part is that anybody with basic computer skills can generate deepfakes to a certain extent. However, generating very high-quality deepfakes needs advanced technology like generative adversarial networks (GANs) \cite{mitra2021easydeep}.

\begin{figure}[htbp]
	\centering
	\includegraphics[width=0.80\linewidth]{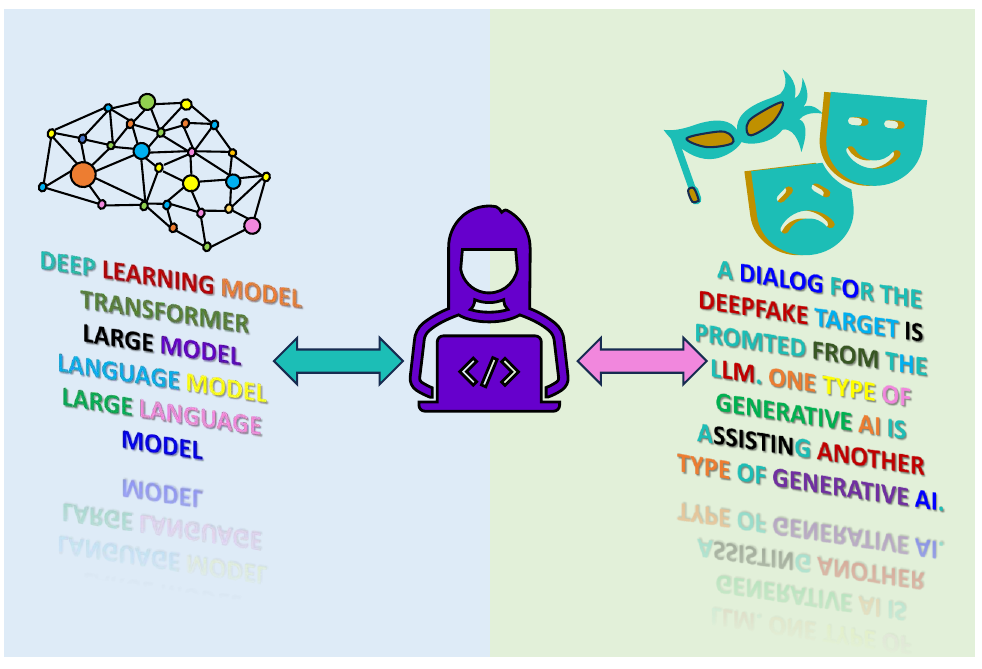}
	\caption{Relation between ChatGPT and Deepfake}
	\label{fig:fig_1}
\end{figure}

This article discusses how generative AI chatbots powered by these large language models can impact deepfake technology (Fig. \ref{fig:fig_1}). In the succeeding sections, we described what a deepfake is, how it is created, what a large language model is, and finally how the AI chatbots can change  deepfake technology.

The rest of this article is organized in the following manner:
The concept of a deepfake is introduced in Section \ref{Sec:Deepfake}. Details of the creation of deepfake videos are presented in Section \ref{Sec:Deepfake_Videos}. Section \ref{Sec:LLM} introduces Large Language Models (LLMs). The roles of ChatGPT for the creation of deepfakes is the scope of Section \ref{Sec:Role_of_ChatGPT}.
Section \ref{Sec:Harnessing_Efforts} outlines some thoughts on efforts to reduce deepfakes.
Conclusions are discussed in Section \ref{Sec:Conclusion}.

\section{What are Deepfakes?}
\label{Sec:Deepfake}

\begin{figure}[h]
	\centering
	\includegraphics[width=0.60\linewidth]{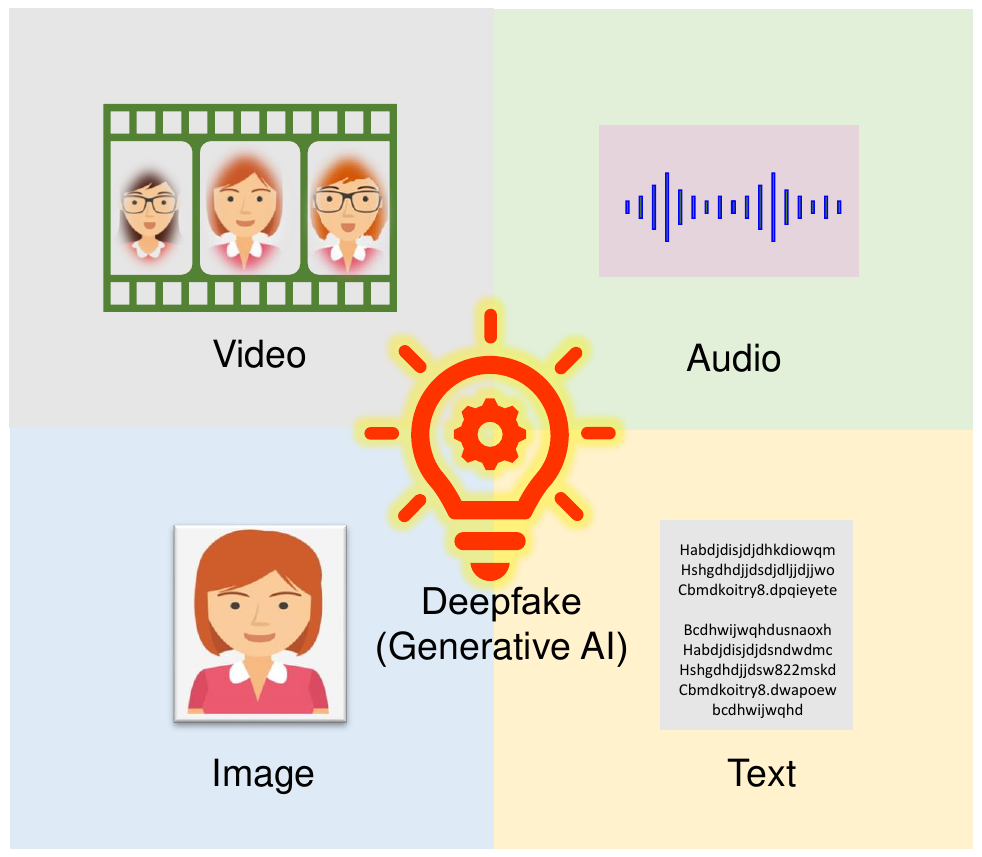}
	\caption{Types of Deepfake.}
	\label{fig:g_ai}
\end{figure}

The term deepfake originated in 2017 by a Reddit user who used deep learning technologies (``deep'') to create fake videos (``fake'') \cite{mitra2020novel}. Deepfake technology allows people to swap faces in videos and images, change voices, and alter texts in documents. Fig. \ref{fig:g_ai} shows different types of deepfakes. They quickly proliferated and have now reached a ``mature'' state. They take digital forgery to a new level. Though graphic-based and content-changing alterations are still common in multimedia forgeries, deep learning technologies have made the forgery much faster, simpler, and more realistic. Earlier generations of deepfakes used to have unnatural movements, artificial facial reactions, sudden shifts in audio quality, mismatches in color and lighting, etc. However, the new generation of deepfakes is sophisticated, and nobody can recognize them with bare eyes. Special deep learning-based methods are required for that.

This technology has spread disinformation, rumors, and propaganda, provoked political strife, blackmailed individuals, and threatened democracy. The capabilities of this technology to manipulate reality have gone too far. This paradigm-shifting technological revolution is altering the truth and creates a zero-trust society insidiously. 



\section{How are Deepfake Videos Created?}
\label{Sec:Deepfake_Videos}

\begin{figure}[h]
	\centering
	\begin{subfigure}[b]{0.5\textwidth}
		\centering
		\includegraphics[width=0.8\linewidth]{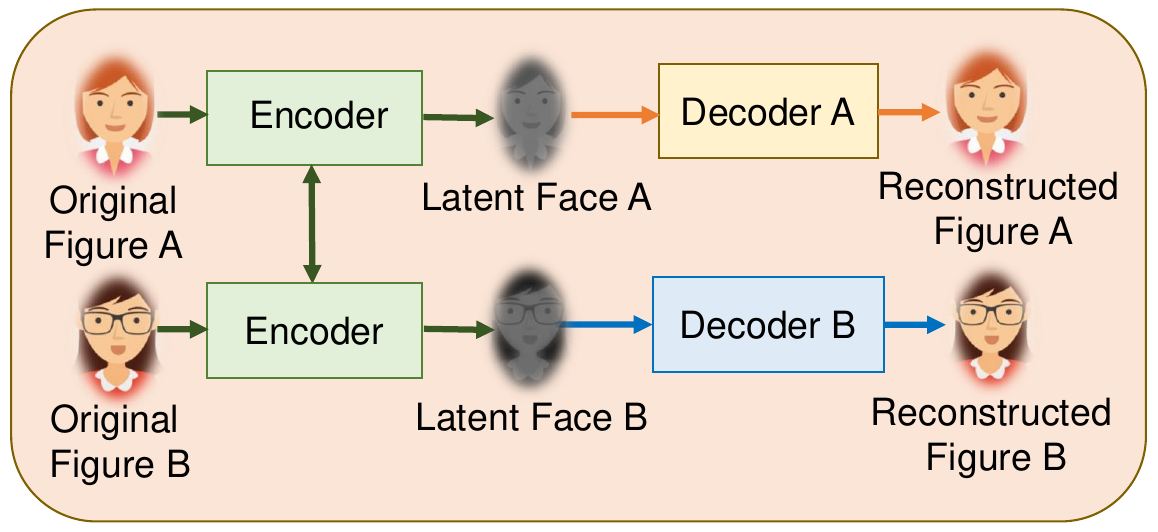}
		\caption{Training of Autoencoders to Create Deepfake}
		\label{fig:df_ae_tr}
	\end{subfigure}
	\hfill
	\begin{subfigure}[b]{.5\textwidth}
		\centering
		\includegraphics[width=0.8\linewidth]{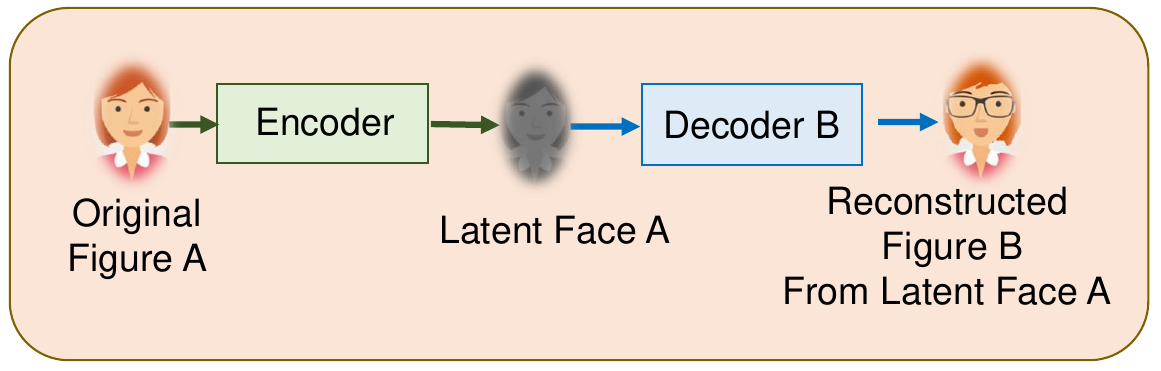}
		\caption{Deepfake Creation using Autoencoder}
		\label{fig:df_ae_ts}
	\end{subfigure}
	\hfill
	\begin{subfigure}[b]{.5\textwidth}
		\centering
		\includegraphics[width=0.8\linewidth]{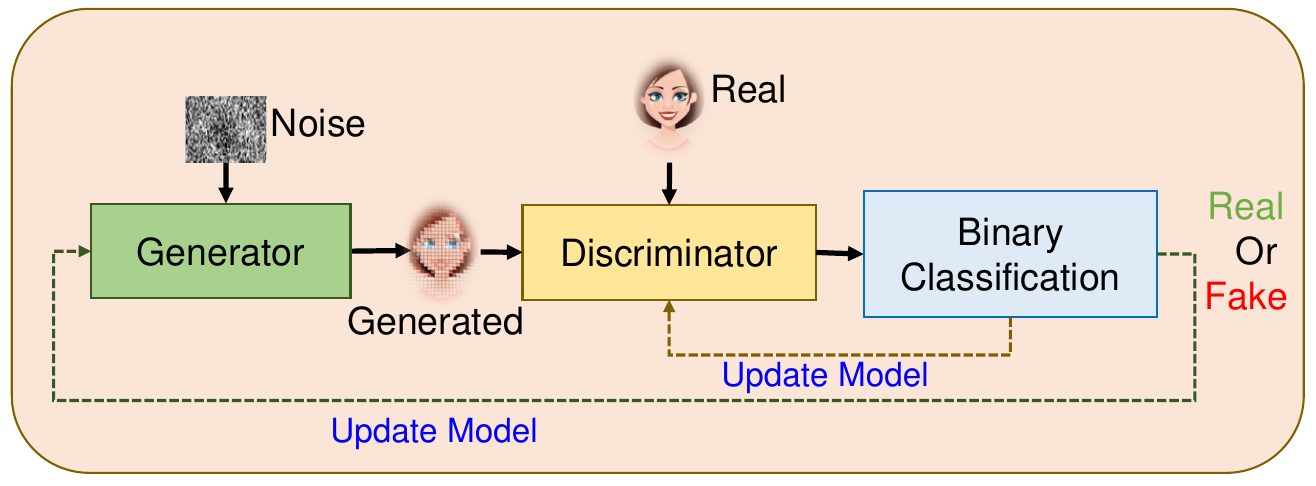}
		\caption{Training of GAN to Create Deepfake}
		\label{fig:df_gan}
	\end{subfigure}
	\caption{Deepfake Creation using Autoencoder and GAN \cite{mitra2021machine}.}
	\label{fig:three_graphs}
\end{figure}

Deepfakes are generated using deep learning models and supervised learning. Different types of deep learning models and different training data are used for the four types of deepfakes. Two key generative neural network architectures for creating deepfake videos are variational autoencoders \cite{kingma2019introduction} and Generative Adversarial Networks (GANs) \cite{goodfellow2020generative}. Facial recognition commonly employs a variational autoencoder to learn latent representations. Autoencoders compress and encode input data (image) to a lower-dimensional latent space, then reconstruct it to provide output data (image). The latent representation comprises all these essential facts that the autoencoder uses to create a more adaptable model that permits ``face swap'' using common features. Figs. \ref{fig:df_ae_tr} and \ref{fig:df_ae_ts} show the training of the autoencoder and the generation of deepfakes using the autoencoder.

Deepfakes use GANs for realism. A ``generator'' creates new images using the latent representation of the source image, and a ``discriminator'' assesses their quality (Fig. \ref{fig:df_gan}). If the generator's image fails the discriminator, it is ordered to create more until one ``fools''the discriminator.

Although it can be hard to make quality deepfake content on a standard personal computer, a plethora of online tools are available that can aid users in this endeavor, e.g., 
FaceApp, TalkingFaces, DeepBrain, Reface, FaceSwap, MorphMe, DeepSwap, DeepFaceLab, etc. ChatGPT integrates with DeepBrain to create real-time conversational deepfakes with custom virtual avatars.

\section{What are Large Language Models?}
\label{Sec:LLM}

Large language models (LLMs) are a type of deep learning model that can perform a variety of natural language processing tasks. In 1967, the world's first chatbot, Eliza, introduced the initial concept of LLMs. Almost three decades later, Long Short Term Memory networks (LSTMs) \cite{6795963} were introduced, which can generate short phrases.

In 2017, the launch of Transfomer (the large language model, ``Attention Is All You Need'') by Google \cite{vaswani2017attention} revolutionized the generative text industry. In 2019, LLMs were able to generate contextual responses and tasks with the launch of Google's Bidirectional Encoder Representations from Transformer (BERT).

In a similar time frame, OpenAI was also working on their LLMs. OpenAI's ChatGPT model began its journey with the release of the GPT-1 model in 2018, which was not made available to the public. An improved version, GPT-2, was published in November 2019. However, GPT-3 (launched in November 2022) \cite{brown2020language} was the ``pivotal moment'' when people were able to interact with the power of generative AI with the launch of OpenAI's chatbot, ChatGPT, or Chat Generative Pre-trained Transformer (GPT).

ChatGPT is a chatbot based on a large-language model. It has prompted a lot of questions due to its technological capabilities which resemble those of humans. ChatGPT is an application of generative AI that uses user input to generate responses in images, videos, and texts. It is even capable of using voice to engage in a lucid conversation. People use it for a variety of purposes, such as:
\begin{itemize}
	\item Generating software codes and checking them for bugs.
	\item Drafting emails.
	\item Summarizing texts.
	\item Writing social network posts.
	\item Scripting conversations.
	\item In general, creating any type of textual content.
\end{itemize}

The original ChatGPT used the GPT-3 model. The recent ChatGPT uses an upgraded version, GPT-3.5. Up to this generation, the LLM accepts only textual input. However, ChatGPT Plus (paid version) and the enterprise version, which use the new GPT-4 LLM \cite{achiam2023gpt}, can accept both text and images as input.

On the other hand, Google's Bard now uses its own model, PaLM-2 \cite{anil2023palm}; earlier, it used to have LaMDA \cite{thoppilan2022lamda}. The specialty of PaLM-2 is that it can summarize a page from a URL. Bing uses the Prometheus model, an ensemble of GPT-4, Bin Orchestrator, and the original Bing search.

Another strong contender in this field is the new LLaMa-2 model \cite{touvron2023llama} by Meta. The size of the LLaMa models (LLaMA-2 and LLaMA) is much smaller compared to the other LLMs. As a result, training LLaMa is much easier and more cost-effective as it requires less computing power. There are other alternatives for other text generators on the market, such as AI-Writer, Deep Write, Jasper, CopySmith, etc., and other coding alternatives like AlphaCode, Amazon CodeWhisper, CodeWp, Seek, Tabine, etc.

\begin{figure}[thbp]
	\centering
	\begin{subfigure}[b]{0.5\textwidth}
		\centering
		\includegraphics[width=0.95\linewidth]{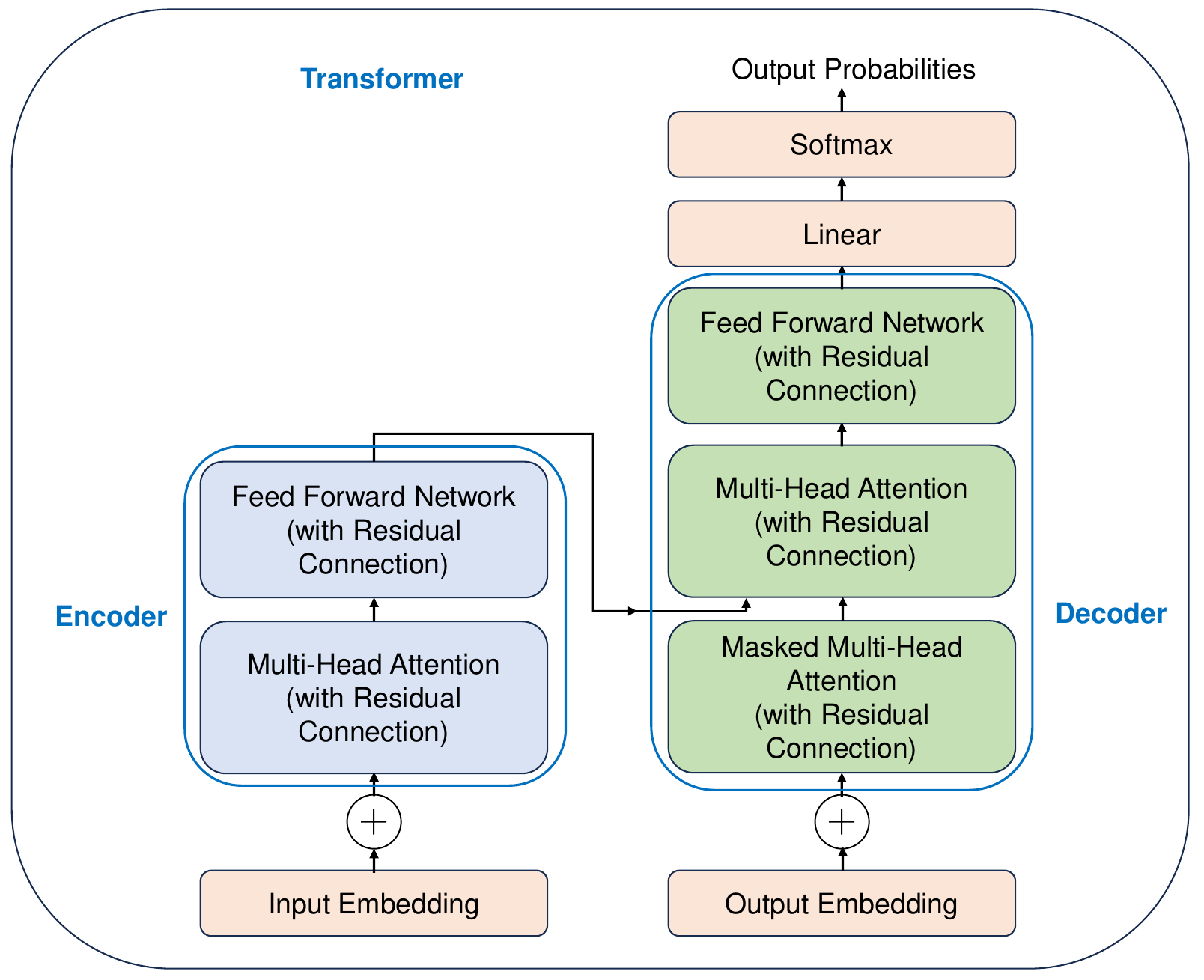}
		\caption{Architecture of Transformer Model \cite{vaswani2017attention}.}
		\label{fig:tr}
	\end{subfigure}
	\hfill
	\begin{subfigure}[b]{.5\textwidth}
		\centering
		\includegraphics[width=0.95\linewidth]{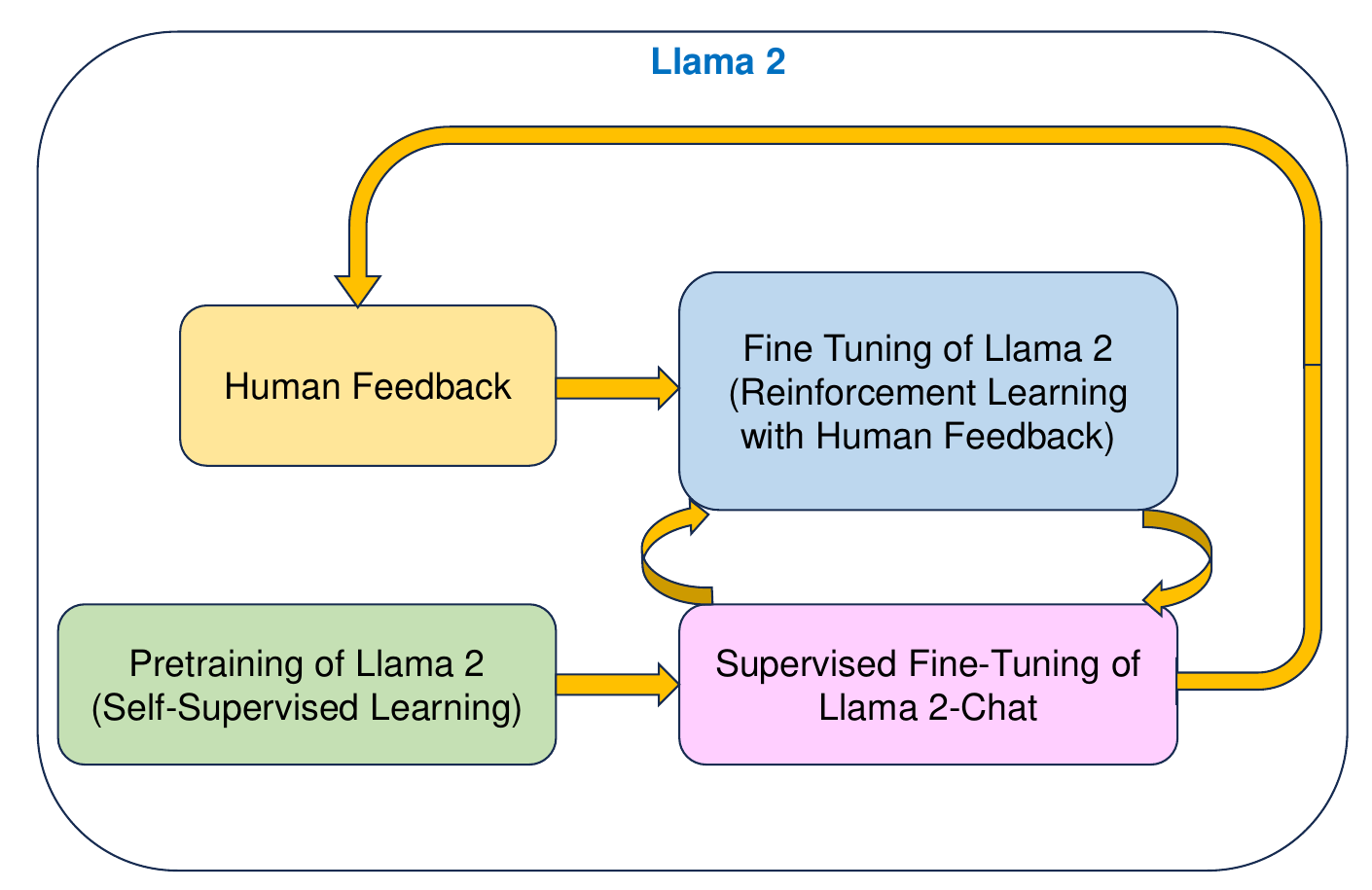}
		\caption{Architecture of Llama 2 \cite{touvron2023llama}.}
		\label{fig:lm}
	\end{subfigure}
	\hfill
	\begin{subfigure}[b]{.5\textwidth}
		\centering
		\includegraphics[width=0.95\linewidth]{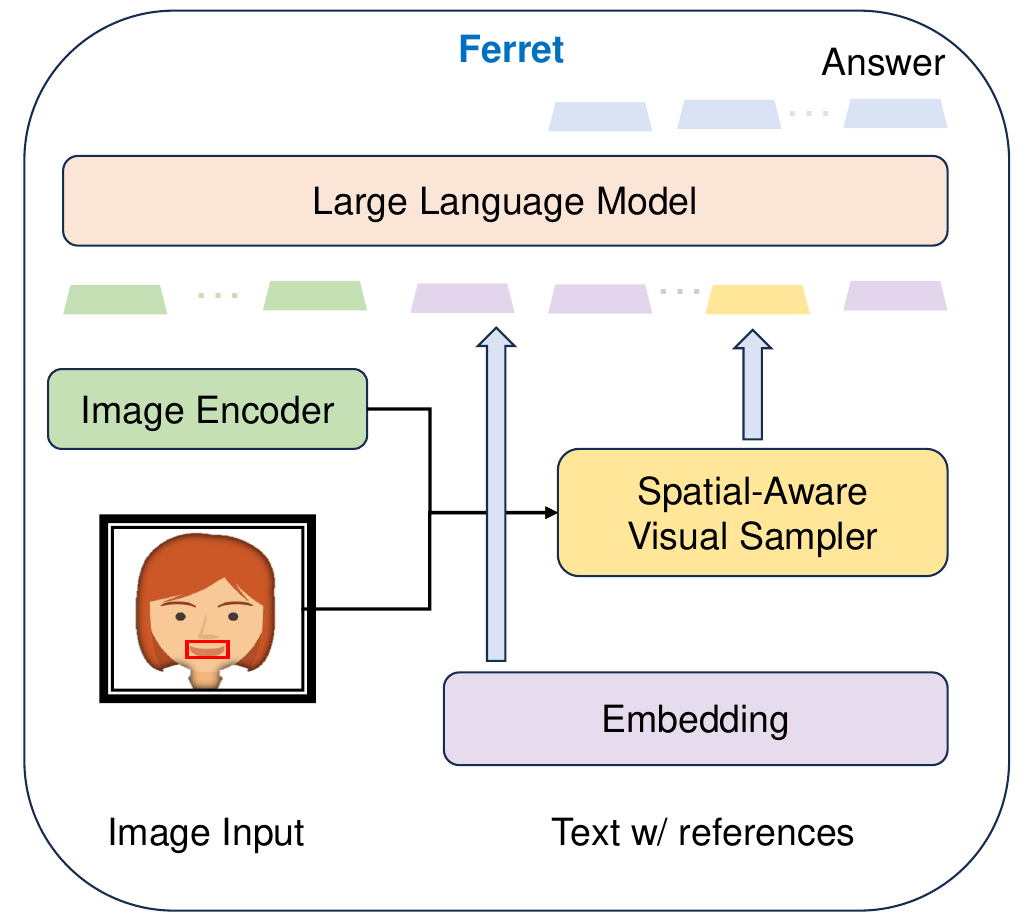}
		\caption{Architecture of Ferret \cite{you2023ferret}.}
		\label{fig:fr}
	\end{subfigure}
	\caption{Architectures of Different LLMs.}
	\label{fig:llm}
\end{figure}

Apple, Inc. is not behind in this game. In October 2023, Apple and Cornell University quietly announced their open-source multi-modal LLM (MLLM), \textit{Ferret} \cite{you2023ferret}. It uses parts of images as the input. Ferret analyzes a region of a picture, detects the elements that potentially answer a question, and creates a bounding box around them. It can then utilize the specified elements to query and respond traditionally.
If a user highlights an animal in a larger image and asks the LLM what it is, it will identify the species. It can then employ other graphic elements to provide further reactions or context for the animal's actions.  
Fig. \ref{fig:llm} shows different LLMs. Fig. \ref{fig:tr} shows the basic \textit{Transformer} architecture. \textit{Bert} is built with the encoder part of the \textit{Transformer } while \textit{GPT 4} is built with the decoder part of it. Fig. \ref{fig:lm} shows the architecture of \textit{Llama 2}. and Fig. \ref{fig:fr} that of \textit{Ferret}.

However, LLMs can create a new level of misinformation, e.g., academic dishonesty, wrong information creation, assisting deepfakes, or creating something from nothing \cite{Shoaib2023DeepfakesMA}.

\section{Role of ChatGPT in Deepfake Creation}
\label{Sec:Role_of_ChatGPT}

Previously, if you wanted to make a convincing dialogue for a deepfake video, you had to compose the lines yourself. These days, it's straightforward to let AI handle everything. You only need to write an outline of the content to obtain a credible dialogue from ChatGPT, Microsoft's Bing chatbot, or any other text generator. They save time and effort by writing the dialogue. Anyone can create dialogue in any language and voice. So, it is not a barrier for anybody to create a deepfake video in another language. It spreads the boundaries of a crime, it can create more engaging dialogue and is cost-effective, as there is no need for professional help.

With OpenAI's new text-writing system, users can now generate lifelike talking heads by simply inputting a prompt and selecting from dozens of avatars and accents created by actors. Various startups like Hour One, Synthesia, and Uneeq are integrating ChatGPT, large language models (LLMs), generative AIs, and other AI with deep learning tools into their platforms to create high-quality synthetic videos without any videographers or high-end video cameras. So, the AI tools create videos that do not exist in real life. In other words, these tools are allowing us to create anything from something else.

Meta is emphasizing open, safe, and responsible generative AI. Purple Llama is the ``umbrella'' project for this effort. They released a set of ``cyber security safety evaluation benchmarks'' \textit{CyberSec Eval} \cite{bhatt2023purple} and a ``safety classifier for input and output filtering'' \textit{Llama Guard} \cite{inan2023llama}. ``Purple'' suggests a collaborative approach for attack and defensive postures.
The discussion will not be complete without mentioning Google's multi-modal model, \textit{Gemini} \cite{team2023gemini}. It can transform any type or mode of input into any type of output. Google claims that it outperforms human experts in Massive Multitask Language Understanding (MMLU) and surpasses all its predecessors from any company. During the short gap between writing and publication of the article, a few more LLMs might emerge.

\section{Harnessing Efforts}
\label{Sec:Harnessing_Efforts}

Political parties, as well as the global community, are keenly cognizant of the tremendous influence that social media exert over the public. World leaders have already made numerous attempts to harness their power. Presently, in light of the emergence of generative AI as a formidable novel instrument, technology behemoths are honing their efforts to thwart its potential propaganda capabilities and voter manipulation.
Alphabet, the parent company of Google, has confirmed that it will restrict the ways in which its chatbot \textit{Bard} and search-generative experience respond to queries concerning the next election \cite{reuters}, adding to an already delicate situation. Meanwhile, Meta, the parent company of Facebook, has prohibited campaigns and other regulated industries from utilizing its generative AI tools to create AI-generated political advertisements. Furthermore, Meta has implemented a policy mandating the disclosure of artificial intelligence (AI) or digital modifications made to political, social, and election-related advertisements on Instagram and Facebook. Meta has taken this action to increase transparency and combat misinformation.
Interestingly, Elon Musk's X reversed its global political ad restriction \cite{forbes}. U.S. candidates and parties can advertise on the site. Governments worldwide are restricting AI-powered political ads. The EU mandates identifying such commercials, including sponsors, costs, and targeted elections.

These efforts are mostly focused on political elections; however, celebrities and common people are also victims of fraud, scams, and blackmail. Deepfake pornographic content targets mostly women to harass  or defame them \cite{hao2021deepfake}. To fight against those situations, academic researchers across the globe are also proposing various methods to address this problem. Unfortunately, these methods alone are insufficient to effectively combat the threat of deepfakes. These methods are mostly based on supervised learning, so when a specific kind of deepfake is detected, the detection method comes in. 

\section{Conclusions}
\label{Sec:Conclusion}

Deepfake and AI chatbots are both the products of generative AI. Both individually threaten the cybersecurity industry, and together they worsen the situation significantly. Both technologies create something that does not exist in reality. Deepfakes are AI-generated fake images or videos. Chatbots make the audio of those videos fluent and lucid. Because of such chatbots, deepfakes can progress to the next stage of perfection.

Although both technologies have several advantages that can help us advance in many sectors, it is unfortunate that, because of their human-imitating power, they are drawing the attention of the global hacker community. As the 2024 U.S. election approaches, the chances of such AI-generated cyber crimes are increasing significantly.

Technology is always one step ahead, and then rules and regulations come. Therefore, it is necessary to establish more stringent laws and regulations related to data privacy and security. Recently, a bill (HF1370) was passed in the Minnesota House stating that nonconsensual sharing of deepfake pornography and political misinformation is a criminal offense. The accused can be charged up to five years of imprisonment and $\$10,000$ in fines. In an example of how politicians do not understand technology, the Senate stalled the bill. India's current prime minister, Mr. Narendra Modi, expressed serious concerns about this technology recently. The IT ministry went one step forward and called this technology a ``new threat to democracy.'' The Indian government has announced a new watchdog website to limit the spread of deepfakes. People can complain about the person or group of people who are making deepfakes. The Indian government is expecting $100\%$ compliance from the social media companies in fighting deepfakes. Sony is developing an in-camera authenticity technology to tackle deepfakes by introducing a digital signature when a picture is taken through Sony cameras, thereby identifying real images. However, smartphone cameras capture most of the images today instead of professional cameras. So, these efforts are not enough. We need more research on deepfake prevention technology and methods that can detect deepfakes in real time and on mobile devices. We believe that more extensive use of digital watermarking, the use of distributed ledgers, and letting people access the metadata of any published videos can stop the spreading of rumors and lies.


\bibliographystyle{IEEEtran}
\bibliography{Bilbiography_Deepfake-ChatGPT}

\begin{thebibliography}{10}
\providecommand{\url}[1]{#1}
\csname url@samestyle\endcsname
\providecommand{\newblock}{\relax}
\providecommand{\bibinfo}[2]{#2}
\providecommand{\BIBentrySTDinterwordspacing}{\spaceskip=0pt\relax}
\providecommand{\BIBentryALTinterwordstretchfactor}{4}
\providecommand{\BIBentryALTinterwordspacing}{\spaceskip=\fontdimen2\font plus
\BIBentryALTinterwordstretchfactor\fontdimen3\font minus
  \fontdimen4\font\relax}
\providecommand{\BIBforeignlanguage}[2]{{%
\expandafter\ifx\csname l@#1\endcsname\relax
\typeout{** WARNING: IEEEtran.bst: No hyphenation pattern has been}%
\typeout{** loaded for the language `#1'. Using the pattern for}%
\typeout{** the default language instead.}%
\else
\language=\csname l@#1\endcsname
\fi
#2}}
\providecommand{\BIBdecl}{\relax}
\BIBdecl

\bibitem{mitra2021machine}
A.~Mitra, S.~P. Mohanty, P.~Corcoran, and E.~Kougianos, ``A machine learning
  based approach for deepfake detection in social media through key video frame
  extraction,'' \emph{SN Computer Science}, vol.~2, pp. 1--18, 2021.

\bibitem{Shoaib2023DeepfakesMA}
\BIBentryALTinterwordspacing
M.~R. Shoaib, Z.~Wang, M.~T. Ahvanooey, and J.~Zhao, ``Deepfakes,
  misinformation, and disinformation in the era of frontier ai, generative ai,
  and large ai models,'' in \emph{2023 International Conference on Computer and
  Applications (ICCA)}, 2023, pp. 1--7. [Online]. Available:
  \url{https://api.semanticscholar.org/CorpusID:265499108}
\BIBentrySTDinterwordspacing

\bibitem{IEEE_Spectrum}
\BIBentryALTinterwordspacing
E.~Strickland, ``Content credentials will fight deepfakes in the 2024
  elections,'' \emph{IEEE Spectrum}, 2023. [Online]. Available:
  \url{https://spectrum.ieee.org/deepfakes-election}
\BIBentrySTDinterwordspacing

\bibitem{holldeep}
\BIBentryALTinterwordspacing
E.~Harrison, ``Itv’s deep fake neighbour wars trailer is the most disturbing
  thing you’ll watch today,'' Independent, November 16, 2022. [Online].
  Available:
  \url{https://www.independent.co.uk/arts-entertainment/tv/news/deep-fake-neighbour-wars-itv-b2226293.html}
\BIBentrySTDinterwordspacing

\bibitem{mitra2021easydeep}
A.~Mitra, S.~P. Mohanty, P.~Corcoran, and E.~Kougianos, ``Easydeep: An iot
  friendly robust detection method for gan generated deepfake images in social
  media,'' in \emph{Proc. of the 4th IFIP International Internet of Things
  Conference}.\hskip 1em plus 0.5em minus 0.4em\relax Springer, 2021, pp.
  217--236.

\bibitem{mitra2020novel}
A.~\hspace{0.01mm}Mitra, S.~P. Mohanty, P.~Corcoran, and E.~Kougianos, ``A
  novel machine learning based method for deepfake video detection in social
  media,'' in \emph{Proc. of IEEE International Symposium on Smart Electronic
  Systems (iSES)(Formerly iNiS)}.\hskip 1em plus 0.5em minus 0.4em\relax IEEE,
  2020, pp. 91--96.

\bibitem{kingma2019introduction}
D.~P. Kingma, M.~Welling \emph{et~al.}, ``An introduction to variational
  autoencoders,'' \emph{Foundations and Trends{\textregistered} in Machine
  Learning}, vol.~12, no.~4, pp. 307--392, 2019.

\bibitem{goodfellow2020generative}
I.~Goodfellow, J.~Pouget-Abadie, M.~Mirza, B.~Xu, D.~Warde-Farley, S.~Ozair,
  A.~Courville, and Y.~Bengio, ``Generative adversarial networks,''
  \emph{Communications of the ACM}, vol.~63, no.~11, pp. 139--144, 2020.

\bibitem{6795963}
S.~Hochreiter and J.~Schmidhuber, ``Long short-term memory,'' \emph{Neural
  Computation}, vol.~9, no.~8, pp. 1735--1780, 1997.

\bibitem{vaswani2017attention}
A.~Vaswani, N.~Shazeer, N.~Parmar, J.~Uszkoreit, L.~Jones, A.~N. Gomez,
  {\L}.~Kaiser, and I.~Polosukhin, ``Attention is all you need,''
  \emph{Advances in neural information processing systems}, vol.~30, 2017.

\bibitem{brown2020language}
T.~Brown, B.~Mann, N.~Ryder, M.~Subbiah, J.~D. Kaplan, P.~Dhariwal,
  A.~Neelakantan, P.~Shyam, G.~Sastry, A.~Askell \emph{et~al.}, ``Language
  models are few-shot learners,'' \emph{Advances in neural information
  processing systems}, vol.~33, pp. 1877--1901, 2020.

\bibitem{achiam2023gpt}
J.~Achiam, S.~Adler, S.~Agarwal, L.~Ahmad, I.~Akkaya, F.~L. Aleman, D.~Almeida,
  J.~Altenschmidt, S.~Altman, S.~Anadkat \emph{et~al.}, ``Gpt-4 technical
  report,'' \emph{arXiv preprint arXiv:2303.08774}, 2023.

\bibitem{anil2023palm}
R.~Anil, A.~M. Dai, O.~Firat, M.~Johnson, D.~Lepikhin, A.~Passos, S.~Shakeri,
  E.~Taropa, P.~Bailey, Z.~Chen \emph{et~al.}, ``Palm 2 technical report,''
  \emph{arXiv preprint arXiv:2305.10403}, 2023.

\bibitem{thoppilan2022lamda}
R.~Thoppilan, D.~De~Freitas, J.~Hall, N.~Shazeer, A.~Kulshreshtha, H.-T. Cheng,
  A.~Jin, T.~Bos, L.~Baker, Y.~Du \emph{et~al.}, ``Lamda: Language models for
  dialog applications,'' \emph{arXiv preprint arXiv:2201.08239}, 2022.

\bibitem{touvron2023llama}
H.~Touvron and et~al., ``Llama 2: Open foundation and fine-tuned chat models,''
  \emph{arXiv, 2307.09288}, 2023.

\bibitem{you2023ferret}
H.~You, H.~Zhang, Z.~Gan, X.~Du, B.~Zhang, Z.~Wang, L.~Cao, S.-F. Chang, and
  Y.~Yang, ``Ferret: Refer and ground anything anywhere at any granularity,''
  \emph{arXiv, 2310.07704}, 2023.

\bibitem{bhatt2023purple}
M.~Bhatt, S.~Chennabasappa, C.~Nikolaidis, S.~Wan, I.~Evtimov, D.~Gabi,
  D.~Song, F.~Ahmad, C.~Aschermann, L.~Fontana \emph{et~al.}, ``Purple llama
  cyberseceval: A secure coding benchmark for language models,'' \emph{arXiv
  preprint arXiv:2312.04724}, 2023.

\bibitem{inan2023llama}
H.~Inan, K.~Upasani, J.~Chi, R.~Rungta, K.~Iyer, Y.~Mao, M.~Tontchev, Q.~Hu,
  B.~Fuller, D.~Testuggine \emph{et~al.}, ``Llama guard: Llm-based input-output
  safeguard for human-ai conversations,'' \emph{arXiv preprint
  arXiv:2312.06674}, 2023.

\bibitem{team2023gemini}
G.~Team, R.~Anil, S.~Borgeaud, Y.~Wu, J.-B. Alayrac, J.~Yu, R.~Soricut,
  J.~Schalkwyk, A.~M. Dai, A.~Hauth \emph{et~al.}, ``Gemini: a family of highly
  capable multimodal models,'' \emph{arXiv preprint arXiv:2312.11805}, 2023.

\bibitem{reuters}
Reuters, ``Alphabet to limit election queries bard and ai-based search can
  answer,''
  https://www.reuters.com/technology/alphabet-limit-election-queries-bard-ai-based-search-can-answer-2023-12-19/,
  December 19, 2023.

\bibitem{forbes}
E.~Woollacott, ``X lifts ban on political ads,''
  https://www.forbes.com/sites/emmawoollacott/2023/08/30/x-lifts-ban-on-political-ads/?sh=1549fbe2222c,
  August 30, 2023.

\bibitem{hao2021deepfake}
K.~Hao, ``Deepfake porn is ruining women's lives. now the law may finally ban
  it,'' \emph{Technology Review}, vol.~12, p. 2021, 2021.

\end{thebibliography}

\section*{Authors}



\begin{wrapfigure}[10]{l}{0.18\linewidth} 
	\vspace{-\baselineskip}
	\centering
	\includegraphics[height=1.3in,keepaspectratio]{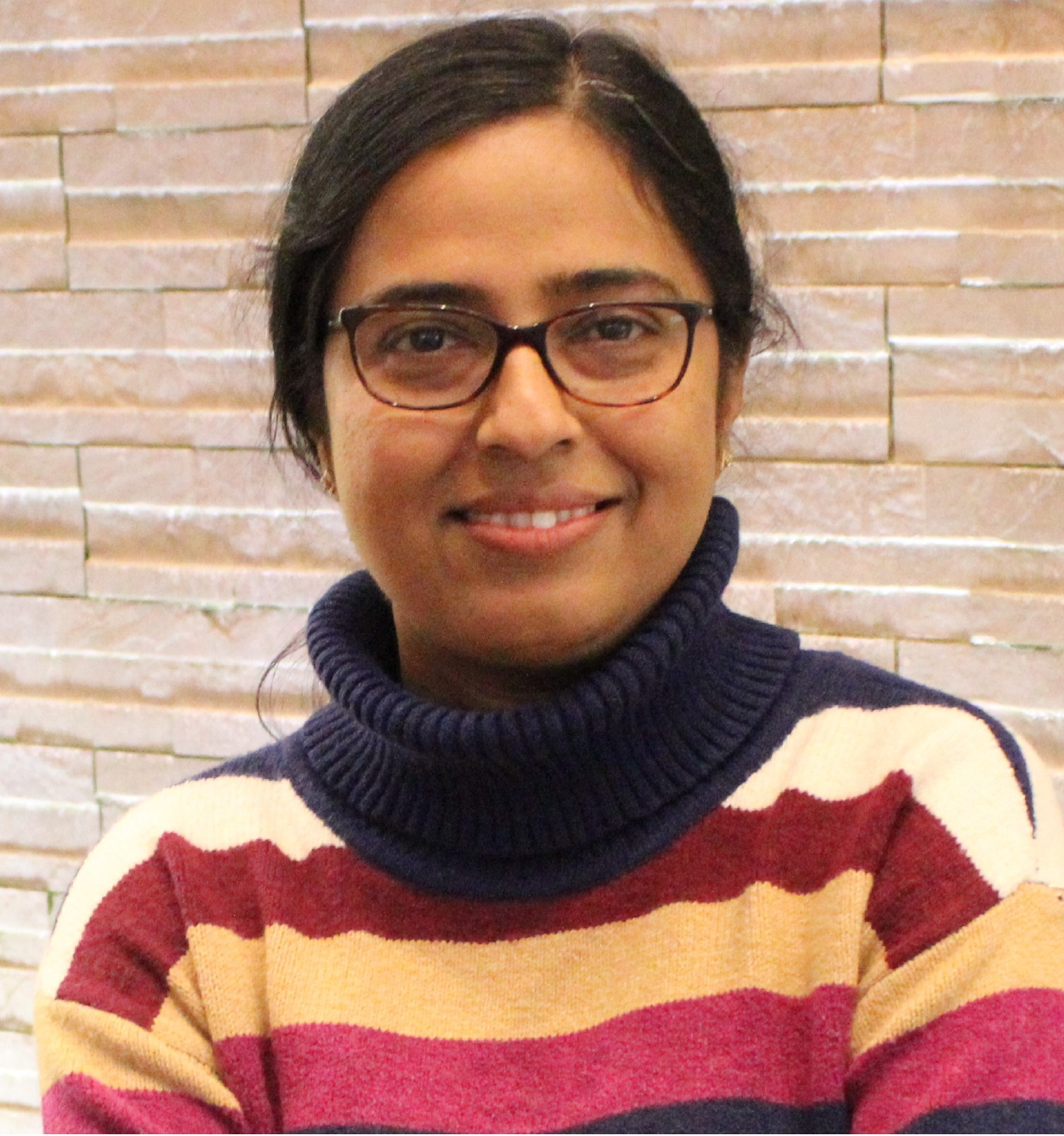}
\end{wrapfigure}
 
\textbf{Alakananda Mitra} is a Research Assistant Professor at the Nebraska Water Center at the Institute of Agriculture and natural Resources, University of Nebraska-Lincoln, Lincoln, NE, USA. She started working as a Visiting Computer Scientist at the USDA-ARS Adaptive Cropping Systems Laboratory at the Beltsville Agricultural Research Center, Beltsville, MD in March 2023. She earned her Ph.D. degree in computer science and engineering from University of North Texas, Denton, TX, USA in 2022. She earned her Bachelor of Science degree (Hons.) in physics from the Presidency College, University of Calcutta, in 2001, and her B. Tech. and M.Tech. degrees in radiophysics and electronics from the Institute ofRadiophysics and Electronics, University of Calcutta, in 2004 and 2006,respectively. Her research interests include application-specific AI/ML/deep learning technologies, computer vision, and edge AI, especially in smart agriculture and multi-media forensics. Currently, she is working on AI-based crop models, tinyMLdevices for plant disease detection, and application of federated learning in smart agriculture. She is also working on a project for developing crop and soil simulation models, graphical user interfaces, databases, and other suitable agro-climatology modeling tools. Dr. Mitra received numerous academic awards, honors, and travel grants throughout her career. During her doctoral research, she received the Outstanding Early-Stage Doctoral Student Award. She also received several Best Paper awards and has one US patent (pending) and one US provisional patent. She is a member of IEEE and American Geophysical Union.

\vspace{1.5cm}
\begin{wrapfigure}{l}{0.18\linewidth} 
	\centering
	\includegraphics[height=1.6in]{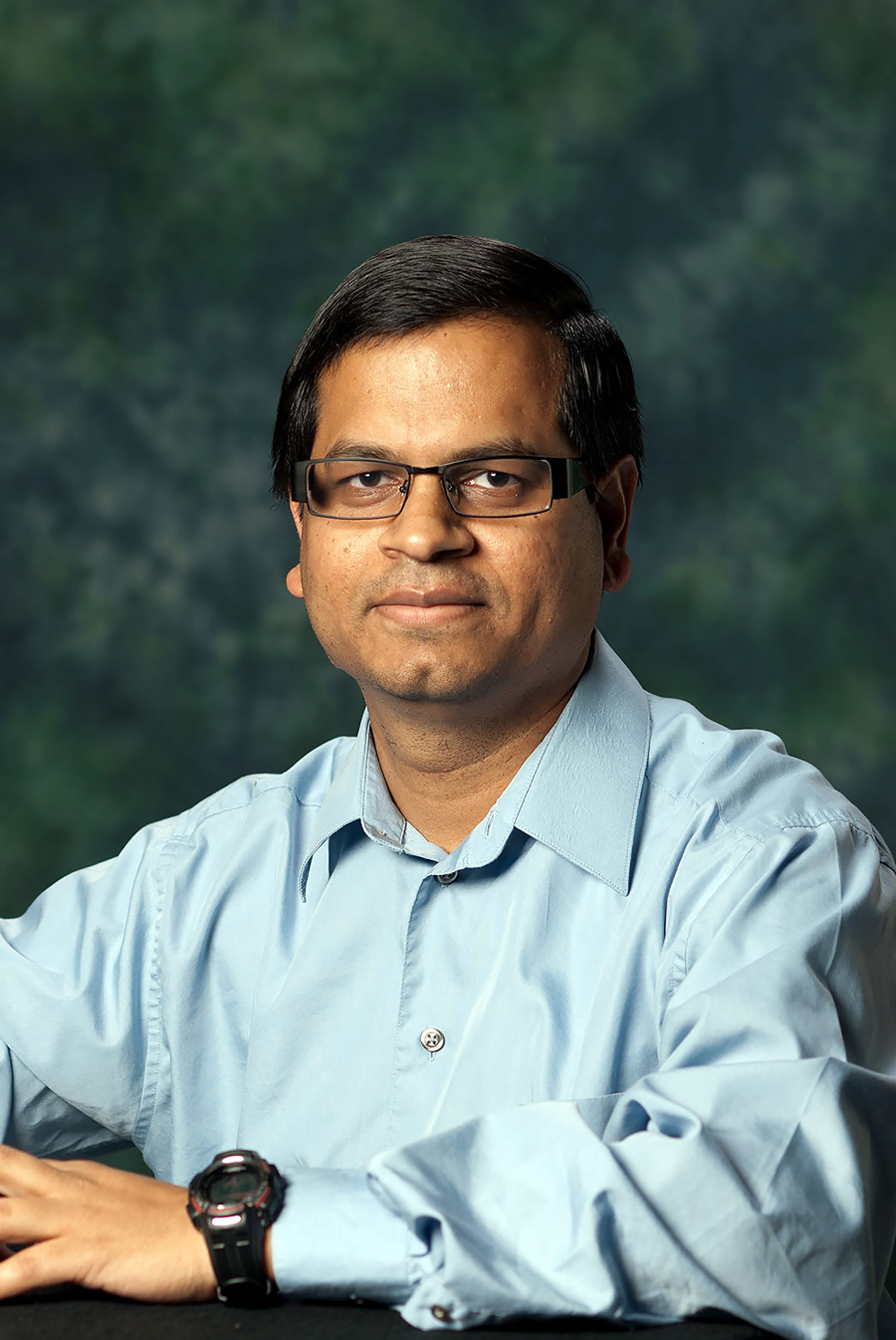}
\end{wrapfigure}
\textbf {Saraju P. Mohanty} received the bachelor’s degree (Honors) in electrical engineering from the Orissa University of Agriculture and Technology, Bhubaneswar, in 1995, the master’s degree in Systems Science and Automation from the Indian Institute of Science, Bengaluru, in 1999, and the Ph.D. degree in Computer Science and Engineering from the University of South Florida, Tampa, in 2003. He is a Professor with the University of North Texas. His research is in ``Smart Electronic Systems’’ which has been funded by National Science Foundations (NSF), Semiconductor Research Corporation (SRC), U.S. Air Force, IUSSTF, and Mission Innovation. He has authored 500 research articles, 5 books, and 10 granted and pending patents. His Google Scholar h-index is 57 and i10-index is 242 with 13,000 citations. He is regarded as a visionary researcher on Smart Cities technology in which his research deals with security and energy aware, and AI/ML-integrated smart components. He introduced the Secure Digital Camera (SDC) in 2004 with built-in security features designed using Hardware Assisted Security (HAS) or Security by Design (SbD) principle. He is widely credited as the designer for the first digital watermarking chip in 2004 and first the low-power digital watermarking chip in 2006. He is a recipient of 18 best paper awards, Fulbright Specialist Award in 2021, IEEE Consumer Electronics Society Outstanding Service Award in 2020, the IEEE-CS-TCVLSI Distinguished Leadership Award in 2018, and the PROSE Award for Best Textbook in Physical Sciences and Mathematics category in 2016. He has delivered 24 keynotes and served on 14 panels at various International Conferences. He has been serving on the editorial board of several peer-reviewed international transactions/journals, including IEEE Transactions on Big Data (TBD), IEEE Transactions on Computer-Aided Design of Integrated Circuits and Systems (TCAD), IEEE Transactions on Consumer Electronics (TCE), and ACM Journal on Emerging Technologies in Computing Systems (JETC). He has been the Editor-in-Chief (EiC) of the IEEE Consumer Electronics Magazine (MCE) during 2016-2021. He served as the Chair of Technical Committee on Very Large Scale Integration (TCVLSI), IEEE Computer Society (IEEE-CS) during 2014-2018 and on the Board of Governors of the IEEE Consumer Electronics Society during 2019-2021. He serves on the steering, organizing, and program committees of several international conferences. He is the steering committee chair/vice-chair for the IEEE International Symposium on Smart Electronic Systems (IEEE-iSES), the IEEE-CS Symposium on VLSI (ISVLSI), and the OITS International Conference on Information Technology (OCIT). He has mentored 3 post-doctoral researchers, and supervised 15 Ph.D. dissertations, 26 M.S. theses, and 21 undergraduate projects.

\vspace{1.5cm}
\begin{wrapfigure}[10]{l}{0.2\linewidth} 
	\vspace{-\baselineskip}
	\includegraphics[height=1.6in,keepaspectratio]{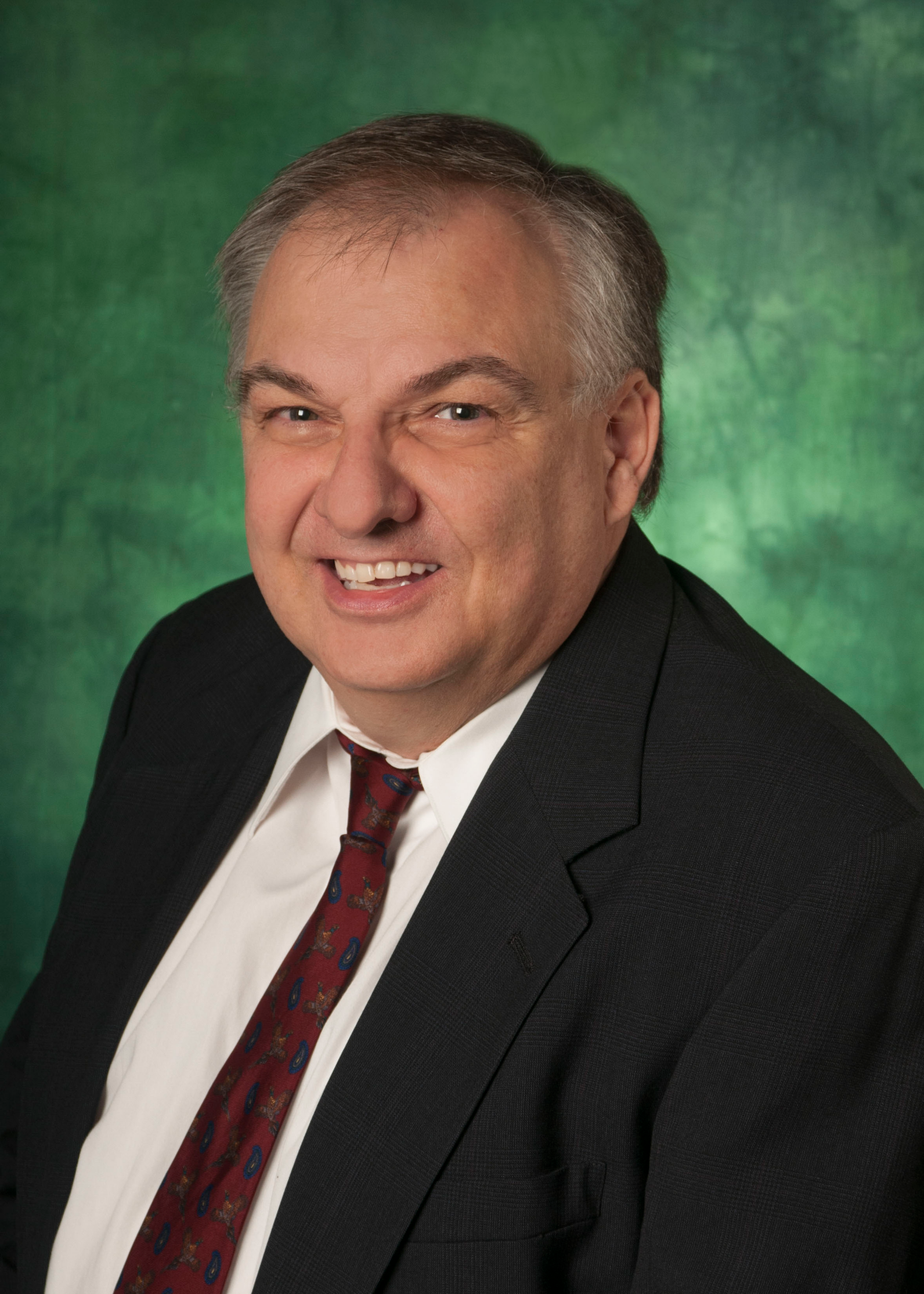}
\end{wrapfigure}
\textbf {Elias Kougianos} received a BSEE from the University of Patras, Greece in 1985 and an MSEE in 1987, an MS in Physics in 1988 and a Ph.D. in EE in 1997, all from Louisiana State University. From 1988 through 1998 he was with Texas Instruments, Inc., in Houston and Dallas, TX. In 1998 he joined Avant! Corp. (now Synopsys) in Phoenix, AZ as a Senior Applications engineer and in 2000 he joined Cadence Design Systems, Inc., in Dallas, TX as a Senior Architect in Analog/Mixed-Signal Custom IC design. He has been at UNT since 2004. He is a Professor in the Department of Electrical Engineering, at the University of North Texas (UNT), Denton, TX. His research interests are in the area of Analog/Mixed-Signal/RF IC design and simulation and in the development of VLSI architectures for multimedia applications. He is an author of over 200 peer-reviewed journal and conference publications.


\end{document}